
\global\newcount\refno
\def\ref#1{\xdef#1{[\the\refno]}
\global\advance\refno by1#1}
\global\refno = 1
\magnification= \magstep 0
\tolerance 10000
\hsize=6 in
\vsize=8.5in
\centerline {\bf FIXED POINTS, SADDLE POINTS AND UNIVERSALITY}
\vskip 1.3 truecm
\centerline{JANOS POLONYI}
\centerline{\it Department of Physics,Louis Pasteur University,Strasbourg,
France}
\centerline{\it Department of Atomic Physics, L. E\"otv\"os University,
Budapest,Hungary}
\vskip 1.3 truecm
\centerline{\bf ABSTRACT}
\vskip 0.8 truecm
\baselineskip 12pt plus 1pt minus 1pt
{\leftskip=3pc\rightskip=3pc\noindent
It is pointed out that the universality might seriously be violated
by models with several fixed points.
\par}
\vskip .6 truecm
\baselineskip 14pt plus 1pt minus 1pt
\noindent{\bf 1. Multiple fixed points}
\medskip
\nobreak
The dependence of the fundamental equations and their parameters on
the observational length scale is described by the renormalization group
equations. In the vicinities of the fixed points the linearized version
of the equations allow us to classify the relevant parameters in a
systematic manner. This classification leads to renormalizable
field theories as the largest class of models to consider.

Unfortunately the
situation is more involved when the description of different interactions
is sought in some unified theory. Consider for the sake of definiteness
a Grand Unified Theory, GUT,
defined at the energy range of $10^{15}$GeV. The interaction vertices of
the lower energy effective theories, such as the electro-weak theory
below 100GeV, QCD around few GeV, nucleon-meson vertices of
of nuclear physics and even the phenomenological coupling constants
of condensed matter physics in the range of MeV and eV, respectively,
are composite operators in terms of the
elementary constituents of the GUT. The true
coupling constant space where the renormalized trajectory generated by the
Wilson-Kadanoff blocking procedure, \ref\wika,
is given includes all of these parameters.
The renormalized trajectory starts at the cut-off, $\Lambda>O(10^{15}GeV)$,
at the GUT ultraviolet fixed point and arrives at the
vicinity of the ultraviolet fixed point of the Standard Model for
$\Lambda=O(10^3GeV)$. In fact, the theory exhibits asymptotic scaling
in this energy range because the super heavy particles of the GUT
scale have already decoupled and the high energy scaling of the
Standard Model is manifest. When the energy is further decresed,
at $\Lambda=O(1GeV)$, the heavy intermediate
bosons and the Higgs particles decouple leaving behind the non-renormalizable
Fermi contact interaction. The QCD evolution equations dominate the
scale dependence in this range.
Thus the renormalized trajectory is in the vicinity
of the QCD fixed point at this energies.

This pattern repeats itself as we move towards lower energies. The
higher energy (effective) theories generate effective vertices.
The renormalizable vertices govern
the renormalized trajectory at the infrared side of the lower
energy fixed point. At the other, ultraviolet side of the
low energy fixed point the relevant coupling constants
are small and the irrelevant ones are dominant. The
non-renormalizable, i.e. irrelevant vertices arising
at the level of the lower energy effective theory
are suppressed according to the decoupling theorem
\ref\apca.
These vertices are nevertheless crucial for they are the seeds of the
"new physics" as seen from the low energy effective theory. In fact, these
are just the coupling constants which grow as we move towards high energy
and deflect the renormalized trajectory from the ultraviolet fixed
point of the low energy effective theory. In our case it is the Fermi
type contact interaction which prevents the physics of few GeV energies
to "fall" into the ultraviolet fixed point of QCD as the energy is raised.
These fixed points which are in the vicinity of the renormalized
trajectory generate cross-over phenomena when their characteristic
scale is reached. It is worthwhile noting that there
are additional fixed points as well which do not belong to either
asymptotical regime of the effective theories.
These correspond to critical phenomena
occurring under special circumstances, such as finite temperature or
densities. Finally there is an infrared fixed point, too.

The universality is the statement that the physics is insensitive to the
initial, UV values of the irrelevant coupling constants. This is certainly
true at the vicinity of the fixed point, more precisely in the infrared end
of the renormalized trajectory within the range of the linearizability.
But in the presence of several fixed points it is no longer clear that
the irrelevant operators of a fixed point remain irrelevant at other
fixed point. Thus we are left only with the "islands" of universality
in the vicinities of different fixed points. The manner
the sets of relevant coupling constants of different fixed
points match is not at all described by universality.

There are two different ways to establish the connection between the
coupling constants of the different fixed points. These will be briefly
described below.
\vskip .4 truecm
\noindent{\bf 2. Scaling at the infrared fixed point}
\medskip
\nobreak
The straightforward procedure is to simply
follow the evolution of the renormalized trajectory.
If there are no further fixed points as we move towards lower energies then
the irrelevance of the coupling constants at the last fixed point
remain valid down to zero energy.
But if there is a chain of fixed point in the vicinity of
the renormalized trajectory as we seen before then
an vertex which was irrelevant at high energy may become relevant at a lower
energy fixed point. In this case the physics at the lower energy
fixed point may in principle depend on the initial value of this
coupling constant. Thus one has to
introduce manifestly non-renormalizable coupling constants at the
high energy fixed point in order to parametrize the physics at low
energies.

In order to understand such a multi-fixed point situation better
one should first consider a theory with two fixed points, one at the
ultraviolet
and the other at the infrared. The infrared fixed point is trivial, i.e.
the its relevant operators are Gaussian in the presence of a mass gap.
In fact, when the block size grows beyond of the Compton wavelength
of the particle the fluctuations are suppressed and the evolution of the
coupling constants slows down. In order to study
the importance of a non-renormalizable coupling constant we need
non-Gaussian relevant operator. This might be found in theories with
massless excitations.

It has been argued in \ref\univ\ by the help of the functional renormalization
group equation \ref\wegner\
that the infrared fixed point of the linear O(N) model in the symmetry broken
phase has nontrivial relevant operators which contain non-renormalizable
vertices. The infrared divergences of the Goldstone modes pile up
and make the beta function negative and divergent for the odd vertices
of the massive particle. How many new important parameters are generated
by this strong coupling vacuum of the Goldstone modes ? Unfortunately
the linearization and the diagonalization of the renormalization
group equation has not yet been done and the actual form of the scaling
operators is not known. But it seems plausible to assume that the
effect the new coupling constants generate is mainly in the deep infrared
limit.
This is because if they start with an O(1) value in the ultraviolet
they are first suppressed as we leave the vicinity of the ultraviolet fixed
point and are amplified only in the asymtotical infrared scaling regime.
The non-perturbative infrared phenomenon which surely takes place in the vacuum
is the spontaneous breakdown of the symmetry and its parameter is
the magnitude of the condensate, $|<\phi_a>|$. Thus it is reasonable to
assume that the only impact these new coupling constants have is that they
influence the magnitude of the condensate. This influence is "hidden"
for the observations made at intermediate energy where these coupling
contants are extremely small. The value of the condensate naturally influences
the dynamics of higher energies indirectly since it determines
the mass gap in the massive sector. Thus the complete set of parameters
of the theory should consists of the usual renormalizable coupling constants
given at the ultraviolet regime and the values of the nontrivial condensates
which are manifest in the infrared. By the help of these
two sets of parameters we can parametrize the physics of the O(N) model.
\vskip .4 truecm
\noindent{\bf 3. Mini-instantons}
\medskip
\nobreak
Another less direct mechanisms which may couple the short and the
long distance regime is operating in theories with
non-homogenous saddle points. As an example consider a
classically scale invariant theory with instantons
\ref\thinst.
The changing the instanton scale leaves the
renormalized action invariant. The actual saddle point expansion
is performed in the bare cut-off theory so it is the bare rather
than the renormalized lagrangian what should be used. The bare regulated
action does have a scale parameter, the cut-off itself. Thus it is
not at all obvious that the action is indeed independent of the instanton
scale.
By adding appropriate irrelevant operators to the lagrangian,
such as higher derivative terms, the action of the saddle point as the
function of the instanton size may develop a
minimum. The size of the saddle point at this minimum is in the order of
magnitude of the only scale parameter, the cut-off. If this minimum
is small enough then these "topological defects"
\ref\topdef\
may form a gas with divergent density.

Qualitatively similar phenomenon, namely the population of the vacuum
by small localized states has already been conjectured in strong coupling
massless QED. There the $e^-e^+$ bound states with the size of the
cut-off condense and change the long distance physics by
the dynamical breaking of the chiral symmetry
\ref\myr.
What is interesting in the mini-instanton gas is that it corresponds to
time dependent coherent states close to the cut-off without relying
on strong coupling constant. In fact, the mini-instantons are present
in asymptotically free theories such as the O(3) model or QCD. The
important question is whether they can modify the scaling at low energy
like the singular $e^-e^+$ pairs of chiral QED.

Though a single mini-instanton can cause no harm to the
renormalized trajectory at a finite i.e. cut-off independent energy,
the gas of these saddle points might well modify the evolution of the
renormalized coupling constants at finite energies. This may happen because
the change of the cut-off generates two qualitatively new effects
in the framework of the dilute instanton gas approximation.
A tree-level cut-off dependence arises from the saddle-point action.
The one-loop dependence comes from the small fluctuation
determinant. When the instantons are well inside of the infrared regime
compared to the cut-off then this dependence agrees with the usual
renormalized trajectory. But the mini-instantons remain in the vicinity of
the cut-off for a large part of the renormalized trajectory and they
continue to modify the trajectory.

The irrelevant terms of the action come from the next fixed point
in the ultraviolet direction as explained above. The
higher energy theory cuts off the low energy effective model and provides
the irrelevant operators which appear as regulator in the low energy
theory.
The dimension of the irrelevant coupling constants is given by the cut-off,
i.e. the scale of the higher energy fixed point. The irrelevant
operators represent the W and Z exchanges in QCD. Thus the mini-instantons
saturate the path integral by the fluctuations of this higher energy
fixed point. If their influence remain finite at low energy they represent
a non-perturbative bridge between neighboring fixed points.

The ratio of the partition function in the one instanton and the flat
sector, $Z_1/Z_0$, was computed in the two dimensional sigma model in
the presence of higher order derivatives in
\ref\mini.
If the higher order derivative terms are really irrelevant then this
infrared quantity should be insensitive to the choice of
the new coupling constants at the cut-off. This was not the case, $Z_1/Z_0$
displayed a clear dependence on the higher order terms in the action.
Let us impose the renormalization condition that $Z_1/Z_0$ is cut-off
independent. Then by assuming that the new coupling constants remain
cut-off independent along the
renormalized trajectory one can obtain the beta function for the
usual coupling constant of the sigma model. The resulting beta function
is negative and asymptotic freedom is preserved but its first coefficient
which is given by the tree level mini-instanton action is not
universal. For sufficiently large values of the higher
derivative operators the mini-instanton action may even become negative.
In this case the vacuum is a densely packed instanton-anti instanton
gas, similarly an anti-ferromagnetic system. Saddle point expansion
is unreliable and the asymptotic freedom is not guaranteed any more
in this phase.

The phenomenon found in the sigma model is likely to be present in QCD, too.
In the framework of the saddle point expansion one expects three new
relevant parameters. These are the size of the min-instanton, its
action and the second derivative of the action with respect to the
instanton size. These parameters are determined at the cut-off by
the dynamics of the higher fixed point. Their appearance in the
beta function at finite energies represent the coupling between
different fixed points.

Naturally one has to be careful and the final conclusion should be
drawn only after computing Green functions in the infrared regime
on the mini-instanton background. This is a formidable task
in an asymptotically free model.
\vskip .4 truecm
\noindent{\bf 4. Conclusions}
\medskip
\nobreak
The nontrivial interplay of different fixed point of the renormalization
group was briefly discussed in this talk from the point of view of the unified
models. There might be another motivation as well to consider this
problem in theories with nontrivial infrared dynamics.
The taking into account such
effect may provide a clue to follow the renormalized trajectory to
regions where the usual perturbation expansion fails. Consider for example QCD
\ref\paris.
The gauge coupling constant blows up according to the
one-loop renormalization group equation at the cross-over separating
the ultraviolet and the infrared scaling regime. The Wilson-Kadanoff
renormalization group procedure suggests the inclusion of all
coupling constant which might be relevant for the evolution into the action.
This means the inclusion of the hadronic effective vertices together
with the gauge coupling constant. This would be double counting in the
hamiltonian formalism where we would use hadrons and quark-gluon states
simultaneously. But this is not the case with the path integral and
we should be able to pass the cross-over regime with this extended QCD
without seeing diverging coupling constants. The onset of the hadronic
world as we enter into the infrared scaling regime would be achieved
by the hadronic vertices and the gauge coupling constant could remain
finite.
\vskip .4 truecm
\centerline{\bf ACKNOWLEDGEMENTS}
\medskip
\nobreak
It is pleasure to thank Vincenzo Branchina and Sen-Ben Liao for
their collaboration which led to the results mentioned above.
\vskip .4 truecm
\centerline{\bf REFERENCES}
\medskip
\nobreak
\hang\par\noindent{\wika} K. Wilson and J.
Kogut, {\it Phys. Rep.} {\bf 12C} (1975) 75;
K. Wilson, {\it Rev. Mod. Phys.} {\bf 47} (1975) 773.
\medskip
\hang\par\noindent{\apca} T. Appelquist and Carazzone, {\it Phys. Rev}
{\bf D11} (1975) 2856.
\medskip
\hang\par\noindent{\univ} S. B. Liao, J. Polonyi,
"Renormalization Group and Universality",
hep-th 9403111, submitted to {\it Ann. Phys.}
\medskip
\hang\par\noindent{\wegner} F.J. Wegner and A. Houghton, {\it Phys. Rev}
{\bf A8} (1972) 401.
\medskip
\hang\par\noindent{\thinst} G. 'tHooft, {\it Phys. Rev.} {\bf D 14} (1976)
3432.
\medskip
\hang\par\noindent{\topdef} M. Luscher, {\it Nucl.Phys.} {\bf B200} (1982) 61.
\medskip
\hang\par\noindent{\myr} P.I. Fomin, V.P. Gusynin, V.A. Miranski and
YU.A. Sytenko, {\it Riv. Nuovo Cimento}, {\bf 6}, No. 5 (1983);
C.N. Leung, S.T. Love , William A. Bardeen,
{\it Nucl. Phys.} {\bf B 323} (1989) 493 ;
C.N. Leung, S.T. Love , William A. Bardeen, {\it Nucl. Phys.}
{\bf B 273} (1986) 649.
\medskip
\hang\par\noindent{\mini} V. Branchina, J. Polonyi, "Instantons in cut-off
theories", hep-th 9401079 to appear in {\it Nucl. Phys.}
\medskip
\hang\par\noindent{\paris} J. Polonyi, in the Proceedings of the Workshop on
``QCD Vacuum Structure and Its Applications'', H.M. Fried and
B. M\"uller eds., World Scientific, 1993, p. 3.
\end